\documentclass[pdftex,twocolumn,epjc3]{svjour3}          % twocolumn

\RequirePackage{amsmath}
\RequirePackage{multirow}
\RequirePackage{subfigure}
\RequirePackage{hyperref}
\RequirePackage{amssymb}
\RequirePackage{color}
\RequirePackage[T1]{fontenc}

\smartqed  % flush right qed marks, e.g. at end of proof

\RequirePackage{graphicx}

\allowdisplaybreaks[3]

\journalname{Eur. Phys. J. C}

\begin{document}

\title{Highly excited and exotic fully-strange tetraquark states}

\author{Rui-Rui Dong\thanksref{t1} \and Niu Su\thanksref{t1} \and Hua-Xing Chen\thanksref{e1}}

\thankstext[$\star$]{t1}{These authors equally contribute to this work.}
\thankstext{e1}{e-mail: hxchen@seu.edu.cn}

\institute{School of Physics, Southeast University, Nanjing 210094, China}

\date{Received: date / Accepted: date}

\maketitle

\begin{abstract}
Some hadrons have the exotic quantum numbers that the traditional $\bar q q$ mesons and $qqq$ baryons can not reach, such as $J^{PC} = 0^{--}/0^{+-}/1^{-+}/2^{+-}/3^{-+}/4^{+-}$, etc. We investigate for the first time the exotic quantum number $J^{PC}=4^{+-}$, and study the fully-strange tetraquark states with such an exotic quantum number. We systematically construct all the diquark-antidiquark interpolating currents, and apply the method of QCD sum rules to calculate both the diagonal and off-diagonal correlation functions. The obtained results are used to construct three mixing currents that are nearly non-correlated, and we use one of them to extract the mass of the lowest-lying state to be $2.85^{+0.19}_{-0.22}$~GeV. We apply the Fierz rearrangement to transform this mixing current to be the combination of three meson-meson currents, and the obtained Fierz identity suggests that this state dominantly decays into the $P$-wave $\phi(1020) f_2^\prime(1525)$ channel. This fully-strange tetraquark state of $J^{PC}=4^{+-}$ is a purely exotic hadron to be potentially observed in future particle experiments.
\end{abstract}

\section{Introduction}
\label{sec:introduction}

In the past twenty years many candidates of exotic hadrons were observed in particle experiments, which can not be well explained in the traditional quark model~\cite{pdg}. Most of them still have the ``traditional'' quantum numbers that the traditional $\bar q q$ mesons and $qqq$ baryons can also reach, making them not so easy to be clearly identified as exotic hadrons. However, there are some ``exotic'' quantum numbers that the traditional hadrons can not reach, such as the spin-parity quantum numbers $J^{PC} = 0^{--}/0^{+-}/1^{-+}/2^{+-}/3^{-+}/4^{+-}/\cdots$. The hadrons with such exotic quantum numbers are of particular interests, since they can not be explained as traditional hadrons any more. Their possible interpretations are compact multiquark states~\cite{Chen:2008qw,Chen:2008ne,Zhu:2013sca,Huang:2016rro}, hadronic molecules~\cite{Zhang:2019ykd,Dong:2022cuw,Ji:2022blw}, glueballs~\cite{Morningstar:1999rf,Chen:2005mg,Mathieu:2008me,Meyer:2004gx,Gregory:2012hu,Athenodorou:2020ani,Qiao:2014vva,Pimikov:2017bkk}, and hybrid states~\cite{Frere:1988ac,Page:1998gz,MILC:1997usn,Dudek:2009qf,Dudek:2013yja,Chetyrkin:2000tj,Chen:2010ic,Huang:2016upt,Qiu:2022ktc,Wang:2022sib}, etc.

Among these exotic quantum numbers, the states of $J^{PC} = 1^{-+}$ have been extensively studied in the literature~\cite{Chen:2008qw,Chen:2008ne,Huang:2016rro,Zhang:2019ykd,Dong:2022cuw,Frere:1988ac,Page:1998gz,MILC:1997usn,Dudek:2009qf,Dudek:2013yja,Chetyrkin:2000tj,Chen:2010ic,Huang:2016upt,Qiu:2022ktc,Wang:2022sib}, since they are predicted to be the lightest hybrid states~\cite{Meyer:2015eta}. Up to now there have been four structures observed in experiments with $J^{PC} = 1^{-+}$, including three isovector states $\pi_1(1400)$~\cite{IHEP-Brussels-LosAlamos-AnnecyLAPP:1988iqi}, $\pi_1(1600)$~\cite{E852:1998mbq}, and $\pi_1(2015)$~\cite{E852:2004gpn} as well as one isoscalar state $\eta_1(1855)$~\cite{Ablikim:2022zze}. Besides, the states of $J^{PC} = 0^{--}/0^{+-}/2^{+-}/3^{-+}$ have also been studied to some extent~\cite{Zhu:2013sca,Ji:2022blw,Morningstar:1999rf,Chen:2005mg,Mathieu:2008me,Meyer:2004gx,Gregory:2012hu,Athenodorou:2020ani,Qiao:2014vva,Pimikov:2017bkk}. These theoretical and experimental studies have significantly improved our understanding on the non-perturbative behaviors of the strong interaction in the low energy region. However, there has not been any investigation on the exotic quantum number $J^{PC} = 4^{+-}$ yet.

\begin{figure}[hbtp]
\begin{center}
\subfigure[]{\includegraphics[width=0.18\textwidth]{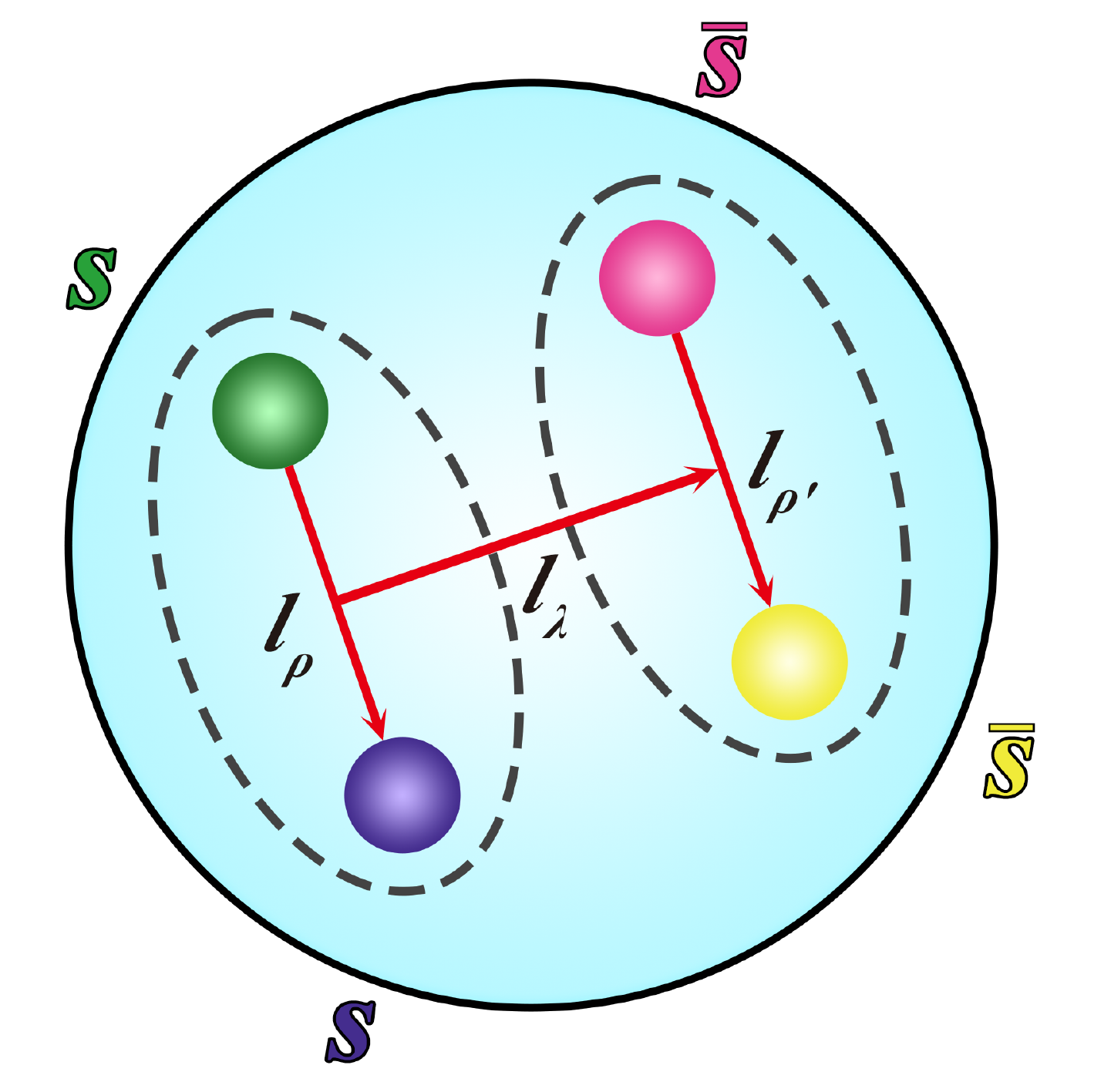}}
~~~~~
\subfigure[]{\includegraphics[width=0.18\textwidth]{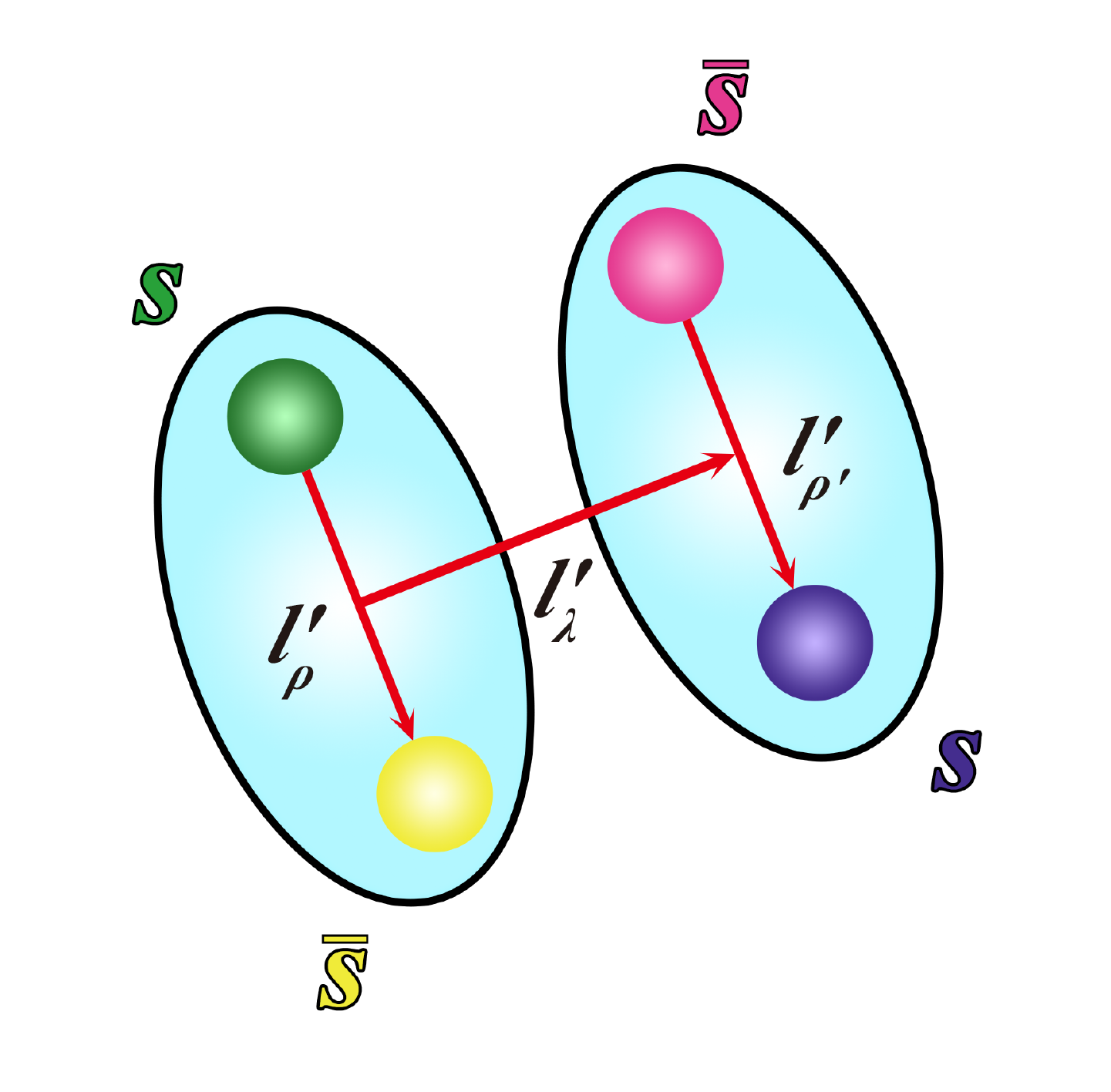}}
\caption{Two configurations for the fully-strange tetraquark states: (a) the diquark-antidiquark system with the internal orbital angular momenta $l_\lambda/l_\rho/l_{\rho^\prime}$ and (b) the meson-meson system with $l_\lambda^\prime/l_\rho^\prime/l_{\rho^\prime}^\prime$.}
\label{fig:orbit}
\end{center}
\end{figure}

In this paper we shall investigate for the first time the exotic quantum number $J^{PC}=4^{+-}$, and study the fully-strange tetraquark states with such an exotic quantum number. We shall work within the diquark-antidiquark picture, and systematically construct all the diquark-antidiquark currents of $J^{PC} = 4^{+-}$, as depicted in Fig.~\ref{fig:orbit}(a). We shall apply the method of QCD sum rules to study these currents as a whole, and extract the mass of the lowest-lying state to be $2.85^{+0.19}_{-0.22}$~GeV.

Besides, we shall also systematically construct all the meson-meson currents of $J^{PC}=4^{+-}$, as depicted in Fig.~\ref{fig:orbit}(b). We shall relate these currents and the diquark-antidiquark currents through the Fierz rearrangement. The obtained Fierz identity suggests that the lowest-lying state dominantly decays into the $P$-wave $\phi(1020) f_2^\prime(1525)$ channel. Accordingly, we propose to search for it in the $X \to \phi(1020) f_2^\prime(1525) \to \phi K \bar K$ decay process. With a large amount of $J/\psi$ sample, the BESIII collaboration are intensively studying the physics happening around here. Such experiments can also be performed by Belle-II, COMPASS, GlueX, and PANDA, etc. Accordingly, this fully-strange tetraquark state of $J^{PC}=4^{+-}$ is a purely exotic hadron to be potentially observed in future particle experiments.

This paper is organized as follows. In Sec.~\ref{sec:current} we systematically construct the local fully-strange tetraquark currents with the exotic quantum number $J^{PC} = 4^{+-}$. We use them to perform QCD sum rule analyses in Sec.~\ref{sec:sumrule}, where we calculate both their diagonal and off-diagonal two-point correlation functions. Based on the obtained results, we use the three single currents to perform numerical analyses in Sec.~\ref{sec:single}, while their mixing currents are investigated in Sec.~\ref{sec:mixing}. Sec.~\ref{sec:summary} is a summary.

\section{Fully-strange tetraquark currents}
\label{sec:current}

As the first step, we construct the local fully-strange tetraquark currents with the exotic quantum number $J^{PC} = 4^{+-}$. This quantum number can not be reached by simply using one quark and one antiquark, and moreover, we need two quarks and two antiquarks together with at least two derivatives to reach such a quantum number.

As depicted in Fig.~\ref{fig:orbit}, there are two possible configurations, the diquark-antidiquark configuration and the meson-meson configuration. When investigating the former configuration, the two covalent derivative operators $D_\alpha (\equiv \partial_\alpha + i g_s A_\alpha)$ and $D_\beta$ can be either inside the diquark/antidiquark field or between them:
\begin{eqnarray}
\nonumber \eta &=& \big[s_a^T C \Gamma_1 {\overset{\leftrightarrow}{D}}_\alpha {\overset{\leftrightarrow}{D}}_\beta s_b \big] (\bar{s}_c \Gamma_2 C \bar{s}_d^T) \pm h.c. \, ,
\\
\nonumber \eta^{\prime} &=& \big[s_a^T C \Gamma_3 {\overset{\leftrightarrow}{D}}_\alpha s_b \big]  \big[\bar{s}_c \Gamma_4 C {\overset{\leftrightarrow}{D}}_\beta \bar{s}_d^T \big] \pm h.c. \, ,
\\
\nonumber \eta^{\prime\prime} &=& \big[\big[s_a^T C \Gamma_5 {\overset{\leftrightarrow}{D}}_\alpha s_b \big] {\overset{\leftrightarrow}{D}}_\beta (\bar{s}_c \Gamma_6 C \bar{s}_d^T)\big] \pm h.c. \, ,
\\
\eta^{\prime\prime\prime} &=& \big[(s_a^T C \Gamma_7 s_b) {\overset{\leftrightarrow}{D}}_\alpha {\overset{\leftrightarrow}{D}}_\beta (\bar{s}_c \Gamma_8 C \bar{s}_d^T)\big] \pm h.c. \, ,
\end{eqnarray}
where $\big[ A {\overset{\leftrightarrow}{D}}_\alpha B \big] \equiv A [D_\alpha B] - [D_\alpha A] B$, $a \cdots d$ are color indices, and $\Gamma_{1\cdots8}$ are Dirac matrices. The internal orbital angular momenta contained in these currents are
\begin{eqnarray}
\nonumber \eta &:& l_\lambda = 0 , \, l_\rho = 2/0 , \, l_{\rho^\prime} = 0/2 ,
\\
\nonumber \eta^{\prime} &:& l_\lambda = 0 , \, l_\rho = 1 , \, l_{\rho^\prime} = 1 ,
\\
\nonumber \eta^{\prime\prime} &:& l_\lambda = 1 , \, l_\rho = 1/0 , \, l_{\rho^\prime} = 0/1 ,
\\
\eta^{\prime\prime\prime} &:& l_\lambda = 2 , \, l_\rho = 0 , \, l_{\rho^\prime} = 0 .
\end{eqnarray}

\begin{figure}[hbtp]
\begin{center}
\includegraphics[width=0.4\textwidth]{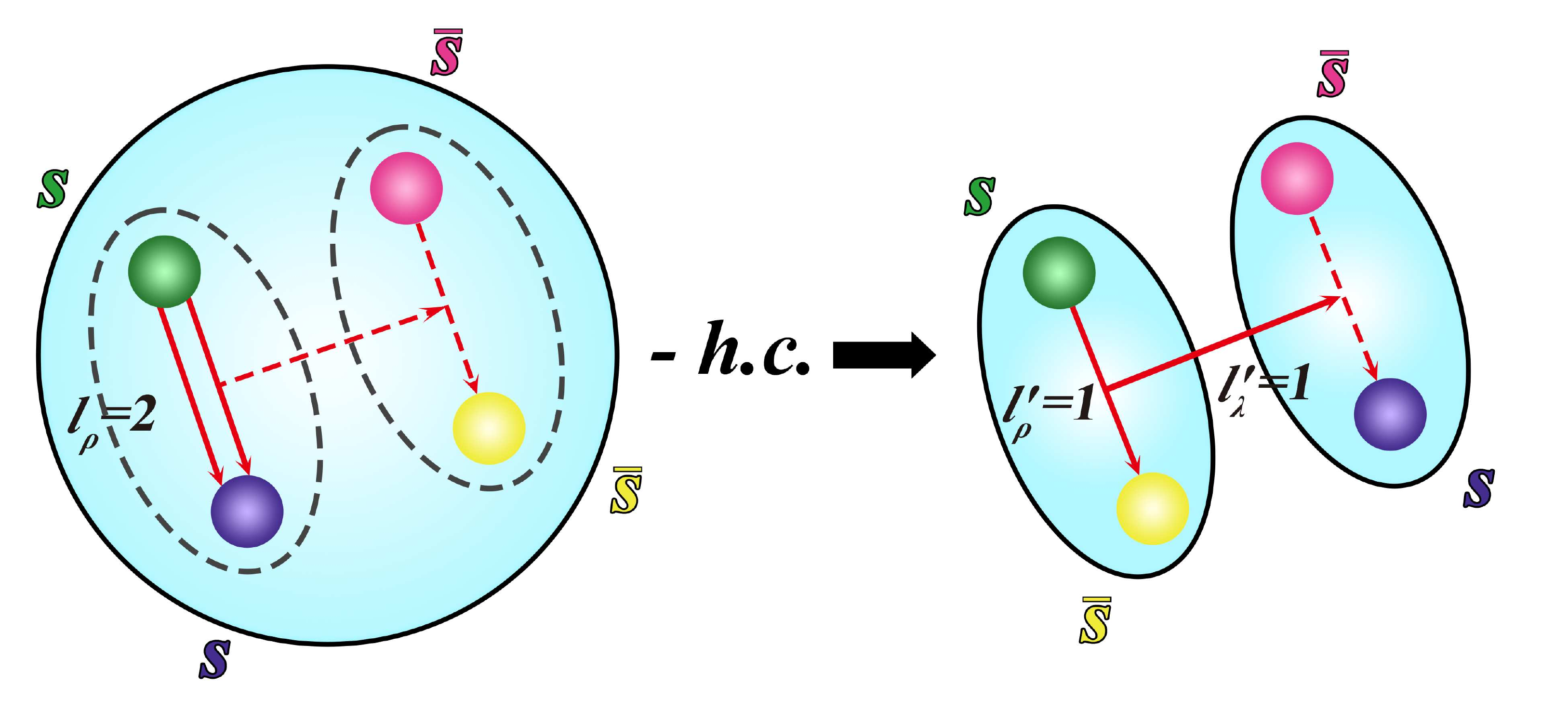}
\caption{Possible internal orbital angular momenta contained in the fully-strange tetraquark currents of $J^{PC} = 4^{+-}$. The Fierz identity given in Eq.~(\ref{eq:fierz}) indicates that the internal orbital angular momenta contained in the diquark-antidiquark system $\{l_\lambda = 0$, $l_\rho = 2/0$, $l_{\rho^\prime} = 0/2\}$ correspond to those contained in the meson-meson system $\{l_\lambda^\prime = 1$, $l_\rho^\prime = 1$, $l_{\rho^\prime}^\prime = 0\}$.}
\label{fig:relation}
\end{center}
\end{figure}

After carefully examining all the possible combinations, we find that only the $\eta$ currents can reach $J^{PC} = 4^{+-}$, as depicted in Fig.~\ref{fig:relation}, while the $\eta^{\prime}/\eta^{\prime\prime}/\eta^{\prime\prime\prime}$ currents can not. Altogether, we can construct three independent diquark-antidiquark currents of $J^{PC} = 4^{+-}$:
\begin{eqnarray}
\nonumber \eta^{1}_{\alpha_1\alpha_2\alpha_3\alpha_4} &=&
\epsilon^{abe} \epsilon^{cde} \times
\\ \nonumber &&
\mathcal{S} \Big\{ \big[s_a^T C \gamma_{\alpha_1} {\overset{\leftrightarrow}{D}}_{\alpha_3}{\overset{\leftrightarrow}{D}}_{\alpha_4} s_b \big] (\bar{s}_c \gamma_{\alpha_2} C \bar{s}_d^T)
\\ \nonumber &&
~ - (s_a^T C \gamma_{\alpha_1} s_b ) \big[\bar{s}_c \gamma_{\alpha_2} C {\overset{\leftrightarrow}{D}}_{\alpha_3} {\overset{\leftrightarrow}{D}}_{\alpha_4} \bar{s}_d^T\big]
\Big\},
\\ \nonumber \eta^{2}_{\alpha_1\alpha_2\alpha_3\alpha_4} &=&
(\delta^{ac} \delta^{bd} + \delta^{ad} \delta^{bc} ) \times
\\ \nonumber &&
\mathcal{S} \Big\{ \big[s_a^T C \gamma_{\alpha_1} \gamma_5 {\overset{\leftrightarrow}{D}}_{\alpha_3}{\overset{\leftrightarrow}{D}}_{\alpha_4} s_b \big] (\bar{s}_c \gamma_{\alpha_2} \gamma_5 C \bar{s}_d^T)
\\ \nonumber &&
~ - (s_a^T C \gamma_{\alpha_1} \gamma_5 s_b ) \big[\bar{s}_c \gamma_{\alpha_2} \gamma_5 C {\overset{\leftrightarrow}{D}}_{\alpha_3} {\overset{\leftrightarrow}{D}}_{\alpha_4} \bar{s}_d^T\big]
\Big\},
\\ \nonumber \eta^{3}_{\alpha_1\alpha_2\alpha_3\alpha_4} &=&
\epsilon^{abe} \epsilon^{cde} g^{\mu\nu} \times
\\ \nonumber &&
\mathcal{S} \Big\{ \big[s_a^T C \sigma_{\alpha_1\mu} {\overset{\leftrightarrow}{D}}_{\alpha_3}{\overset{\leftrightarrow}{D}}_{\alpha_4} s_b \big] (\bar{s}_c \sigma_{\alpha_2\nu} C \bar{s}_d^T)
\\ \nonumber &&
~ - (s_a^T C \sigma_{\alpha_1\mu} s_b ) \big[\bar{s}_c \sigma_{\alpha_2\nu} C {\overset{\leftrightarrow}{D}}_{\alpha_3} {\overset{\leftrightarrow}{D}}_{\alpha_4} \bar{s}_d^T\big]
\Big\}.
\\ \label{def:eta}
\end{eqnarray}
The symbol $\mathcal{S}$ denotes symmetrization and subtracting the trace terms in the set $\{\alpha_1 \cdots \alpha_J\}$. Among these currents, $\eta^{1}_{\cdots}$ and $\eta^{3}_{\cdots}$ have the antisymmetric color structure $[ss]_{\mathbf{\bar 3}_C}[\bar s \bar s]_{\mathbf{3}_C}$, and $\eta^{2}_{\cdots}$ has the symmetric color structure $[ss]_{\mathbf{6}_C}[\bar s \bar s]_{\mathbf{\bar 6}_C}$.

After similarly investigating the meson-meson configuration, we can also construct three independent meson-meson currents of $J^{PC} = 4^{+-}$:
\begin{eqnarray}
\nonumber
\xi^{1}_{\alpha_1\alpha_2\alpha_3\alpha_4} &=&
\mathcal{S} \Big\{
\big[\bar s_a \gamma_{\alpha_1} {\overset{\leftrightarrow}{D}}_{\alpha_3} s_a \big] {\overset{\leftrightarrow}{D}}_{\alpha_4} (\bar{s}_b \gamma_{\alpha_2} s_b)
\Big\} ,
\\ \nonumber \xi^{2}_{\alpha_1\alpha_2\alpha_3\alpha_4} &=&
\mathcal{S} \Big\{
\big[\bar s_a \gamma_{\alpha_1} \gamma_5 {\overset{\leftrightarrow}{D}}_{\alpha_3} s_a \big] {\overset{\leftrightarrow}{D}}_{\alpha_4} (\bar{s}_b \gamma_{\alpha_2} \gamma_5 s_b)
\Big\} ,
\\ \nonumber \xi^{3}_{\alpha_1\alpha_2\alpha_3\alpha_4} &=&
g^{\mu\nu} \mathcal{S} \Big\{
\big[\bar s_a \sigma_{\alpha_1\mu} {\overset{\leftrightarrow}{D}}_{\alpha_3} s_a \big] {\overset{\leftrightarrow}{D}}_{\alpha_4} (\bar{s}_b \sigma_{\alpha_2\nu} s_b)
\Big\} .
\\ \label{def:xi}
\end{eqnarray}
As depicted in Fig.~\ref{fig:relation}, the internal orbital angular momenta contained in these currents are
\begin{eqnarray}
\xi &:& l_\lambda^\prime = 1 , \, l_\rho^\prime = 1 , \, l_{\rho^\prime}^\prime = 0 .
\end{eqnarray}

After applying the Fierz rearrangement, we obtain
\begin{equation}
\left(\begin{array}{c}
\eta^{1}_{\cdots}
\\
\eta^{2}_{\cdots}
\\
\eta^{3}_{\cdots}
\end{array}\right)
=
\left(\begin{array}{ccc}
2 & -2 & -2
\\
-2 & 2 & -2
\\
-4 & -4 & 0
\end{array}\right)
\left(\begin{array}{c}
\xi^{1}_{\cdots}
\\
\xi^{2}_{\cdots}
\\
\xi^{3}_{\cdots}
\end{array}\right)
\, .
\label{eq:fierz}
\end{equation}
This Fierz identity will be used to study the decay behaviors later.

\section{QCD sum rule analysis}
\label{sec:sumrule}

We apply the QCD sum rule method~\cite{Shifman:1978bx,Reinders:1984sr} to study the fully-strange tetraquark currents $\eta^{1,2,3}_{\alpha_1\alpha_2\alpha_3\alpha_4}$ with the exotic quantum number $J^{PC} = 4^{+-}$. This non-perturbative method has been successfully applied to study various conventional and exotic hadrons in the past fifty years~\cite{Nielsen:2009uh}.

We generally assume that the current $\eta^{i}_{\alpha_1\alpha_2\alpha_3\alpha_4}$ ($i=1\cdots3$) couples to the fully-strange tetraquark states $X_n$ ($n=1 \cdots N$) through
\begin{equation}
\langle 0| \eta^i_{\alpha_1\alpha_2\alpha_3\alpha_4} | X_n \rangle = f_{in} \epsilon_{\alpha_1\alpha_2\alpha_3\alpha_4} \, ,
\label{eq:defg}
\end{equation}
where $f_{in}$ is the $3 \times N$ matrix for the decay constants, and $\epsilon_{\alpha_1\alpha_2\alpha_3\alpha_4}$ is the traceless and symmetric polarization tensor satisfying
\begin{equation}
\epsilon_{\alpha_1\alpha_2\alpha_3\alpha_4} \epsilon^*_{\beta_1\beta_2\beta_3\beta_4} = \mathcal{S}^\prime [\tilde g_{\alpha_1 \beta_1} \tilde g_{\alpha_2 \beta_2} \tilde g_{\alpha_3 \beta_3} \tilde g_{\alpha_4 \beta_4}] \, ,
\end{equation}
with $\tilde g_{\mu \nu} = g_{\mu \nu} - q_\mu q_\nu / q^2$. The symbol $\mathcal{S}^\prime$ denotes symmetrization and subtracting the trace
terms in the sets $\{\alpha_1\alpha_2\alpha_3\alpha_4\}$ and $\{\beta_1\beta_2\beta_3\beta_4\}$.

Based on Eq.~(\ref{eq:defg}), we can investigate both the diagonal and off-diagonal correlation functions:
\begin{eqnarray}
\nonumber && \Pi^{ij}_{\alpha_1\alpha_2\alpha_3\alpha_4;\beta_1\beta_2\beta_3\beta_4}(q^2)
\\ \nonumber &\equiv& i \int d^4x e^{iqx} \langle 0 | {\bf T}[ \eta^i_{\alpha_1\alpha_2\alpha_3\alpha_4}(x) { \eta_{\beta_1\beta_2\beta_3\beta_4}^{j,\dagger} } (0)] | 0 \rangle
\\ &=& \Pi_{ij}(q^2) \times \mathcal{S}^\prime [\tilde g_{\alpha_1 \beta_1} \tilde g_{\alpha_2 \beta_2} \tilde g_{\alpha_3 \beta_3} \tilde g_{\alpha_4 \beta_4}] \, .
\label{def:pi}
\end{eqnarray}

At the hadron level we express $\Pi_{ij}(q^2)$ using the dispersion relation as
%
%%%%%%%%%%%%%%%%%%%%%%%%%%%%%%%%%%%%%%%%%%%%%%%%%%%%%%%%%%%%%%%%%%%%%%%%%%%%%%
\begin{equation}
\Pi_{ij}(q^2) = \int^\infty_{s_<}\frac{\rho^{\rm phen}_{ij}(s)}{s-q^2-i\varepsilon}ds \, ,
\end{equation}
%%%%%%%%%%%%%%%%%%%%%%%%%%%%%%%%%%%%%%%%%%%%%%%%%%%%%%%%%%%%%%%%%%%%%%%%%%%%%%
%
with $s_< = 16 m_s^2$ the physical threshold. We parameterize the spectral density $\rho^{\rm phen}_{ij}(s)$ for the states $X_n$ together with a continuum contribution as
%
%%%%%%%%%%%%%%%%%%%%%%%%%%%%%%%%%%%%%%%%%%%%%%%%%%%%%%%%%%%%%%%%%%%%%%%%%%%%%%
\begin{eqnarray}
\nonumber && \rho^{\rm phen}_{ij}(s) \times \mathcal{S}^\prime [\cdots]
\\ \nonumber &\equiv& \sum_n \delta(s-M^2_n) \langle 0| \eta^i_{\cdots} | n \rangle \langle n | {\eta^{j\dagger}_{\cdots}} |0 \rangle + \cdots
\\ &=& \sum_n f_{in}f_{jn} \delta(s-M^2_n) \times \mathcal{S}^\prime [\cdots] + \cdots \, ,
\label{eq:rho}
\end{eqnarray}
with $M_n$ the mass of $X_n$.

\begin{figure}[hbtp]
\begin{center}
\includegraphics[width=0.5\textwidth]{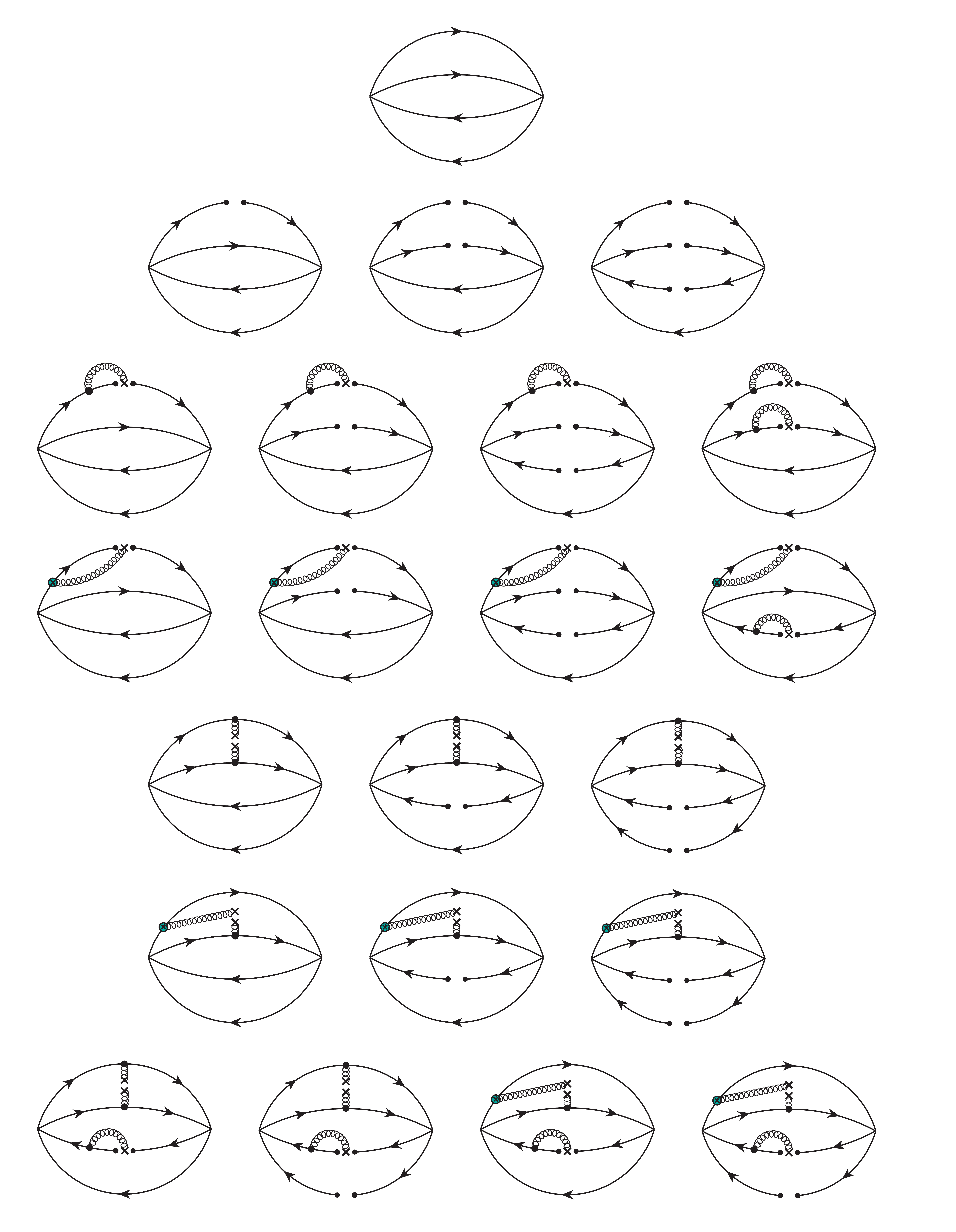}
\caption{Feynman diagrams for the fully-strange tetraquark currents of $J^{PC} = 4^{+-}$. The covariant derivative operator $D_\alpha = \partial_\alpha + i g_s A_\alpha$ contains two terms, and we depict the latter term using a green vertex.}
\label{fig:feynman}
\end{center}
\end{figure}

At the quark-gluon level we calculate $\Pi_{ij}(q^2)$ using the method of operator product expansion (OPE), and extract the OPE spectral density $\rho_{ij}(s) \equiv \rho^{\rm OPE}_{ij}(s)$~\cite{ope}. In the calculations we take into account the Feynman diagrams depicted in Fig.~\ref{fig:feynman}. We consider the perturbative term, the strange quark mass $m_s$, the quark condensate $\langle \bar s s \rangle$, the quark-gluon mixed condensate $\langle g_s \bar s \sigma G s \rangle$, the gluon condensate $\langle g_s^2 GG \rangle$, and their combinations. We calculate all the diagrams proportional to $g_s^{N=0}$ and $g_s^{N=1}$, where we find the $D=6$ term $\langle \bar s s \rangle^2$ and the $D=8$ term $\langle \bar s s \rangle\langle g_s \bar s \sigma G s \rangle$ to be important. We partly calculate the diagrams proportional to $g_s^{N\geq2}$, whose contributions are found to be small. Especially, we have not taken into account the radiative corrections in our QCD sum rule calculations.

Then we perform the Borel transformation at both the hadron and quark-gluon levels. After approximating the continuum using $\rho_{ij}(s)$ above the threshold value $s_0$, we obtain the sum rule equation
%
%%%%%%%%%%%%%%%%%%%%%%%%%%%%%%%%%%%%%%%%%%%%%%%%%%%%%%%%%%%%%%%%%%%%%%%%%%%%%%
\begin{eqnarray}
\nonumber \Pi_{ij}(s_0, M_B^2) &\equiv& \sum_n f_{in}f_{jn} e^{-M_n^2/M_B^2}
\\ &=& \int^{s_0}_{s_<} e^{-s/M_B^2} \rho_{ij}(s) ds \, .
\label{eq:fin}
\end{eqnarray}
%%%%%%%%%%%%%%%%%%%%%%%%%%%%%%%%%%%%%%%%%%%%%%%%%%%%%%%%%%%%%%%%%%%%%%%%%%%%%%
%
We shall investigate it through two steps, the single-channel analysis and the multi-channel analysis, as follows.

\section{Single-channel analysis}
\label{sec:single}

To perform the single-channel analysis, we neglect the off-diagonal correlation functions by setting $\rho_{ij}(s)|_{i \neq j} = 0$ so that only $\rho_{ii}(s) \neq 0$. This assumption means that the three currents $\eta^{1,2,3}_{\alpha_1\alpha_2\alpha_3\alpha_4}$ are ``non-correlated'', and any two of them can not mainly couple to the same state $X$, otherwise,
\begin{eqnarray}
\nonumber && \rho_{ij}(s) \times \mathcal{S}^\prime [\cdots]
\\ \nonumber &\equiv& \sum_n \delta(s-M^2_n) \langle 0| \eta^i_{\cdots} | n \rangle \langle n | {\eta^{j\dagger}_{\cdots}} |0 \rangle + \cdots
\\ \nonumber &\approx& \delta(s-M^2_X) \langle 0| \eta^i_{\cdots} | X \rangle \langle X | {\eta^{j\dagger}_{\cdots}} |0 \rangle + \cdots
\\ &\neq& 0 \, .
\end{eqnarray}
Accordingly, we assume that there are three states $X_{1,2,3}$ corresponding to the three currents $\eta^{1,2,3}_{\alpha_1\alpha_2\alpha_3\alpha_4}$ through
\begin{equation}
\langle 0| \eta^i_{\alpha_1\alpha_2\alpha_3\alpha_4} | X_i \rangle = f_{ii} \epsilon_{\alpha_1\alpha_2\alpha_3\alpha_4} \, .
\end{equation}
After parameterizing the spectral density $\rho_{ii}(s)$ as one pole dominance for the state $X_i$ together with a continuum contribution, Eq.~(\ref{eq:fin}) is simplified to be
\begin{equation}
\Pi_{ii}(s_0, M_B^2) \equiv f_{ii}^2 e^{-M_i^2/M_B^2} = \int^{s_0}_{s_<} e^{-s/M_B^2} \rho_{ii}(s) ds \, ,
\end{equation}
which can be used to calculate $M_i$ through
%
%%%%%%%%%%%%%%%%%%%%%%%%%%%%%%%%%%%%%%%%%%%%%%%%%%%%%%%%%%%%%%%%%%%%%%%%%%%%%%
\begin{equation}
M^2_i(s_0, M_B) = \frac{\int^{s_0}_{s_<} e^{-s/M_B^2} s \rho_{ii}(s) ds}{\int^{s_0}_{s_<} e^{-s/M_B^2} \rho_{ii}(s) ds} \, .
\label{eq:LSR}
\end{equation}
%%%%%%%%%%%%%%%%%%%%%%%%%%%%%%%%%%%%%%%%%%%%%%%%%%%%%%%%%%%%%%%%%%%%%%%%%%%%%%
%

We use the spectral density $\rho_{11}(s)$ extracted from the current $\eta^1_{\alpha_1\alpha_2\alpha_3\alpha_4}$ as an example to perform the numerical analysis. We take the following values for various QCD parameters~\cite{pdg,Yang:1993bp,Narison:2002pw,Gimenez:2005nt,Jamin:2002ev,Ioffe:2002be,Ovchinnikov:1988gk,Ellis:1996xc}:
\begin{eqnarray}
\nonumber m_{s}(2~{\rm GeV}) &=&93_{-~5}^{+11} {\rm~MeV} \, ,
\\ \nonumber
\left\langle g_{s}^{2} G G\right\rangle &=&(0.48 \pm 0.14) {\rm~GeV}^{4} \, ,
\\
\langle\bar ss\rangle &=& -(0.8\pm 0.1)\times(0.240 \mbox{ GeV})^3\, ,
\\ \nonumber
\left\langle g_{s} \bar{s} \sigma G s\right\rangle &=&-M_{0}^{2} \times\langle\bar{s} s\rangle \, ,
\\ \nonumber
M_{0}^{2} &=&(0.8 \pm 0.2) {\rm~GeV}^{2} \, .
\label{eq:condensates}
\end{eqnarray}

As shown in Eq.~(\ref{eq:LSR}), the mass $M_1$ of the state $X_1$ depends on two free parameters, the Borel mass $M_B$ and the threshold value $s_0$. We investigate three aspects to find their proper working regions: a) the convergence of OPE, b) the sufficient amount of the pole contribution, and c) the mass dependence on these two parameters.

%
%%%%%%%%%%%%%%%%%%%%%%%%%%%%%%%%%%%%%%%%%%%%%%%%%%%%%%%%%%%%%%%%%%%%%%%%%%%%%%
\begin{figure}[hbt]
\begin{center}
\includegraphics[width=0.47\textwidth]{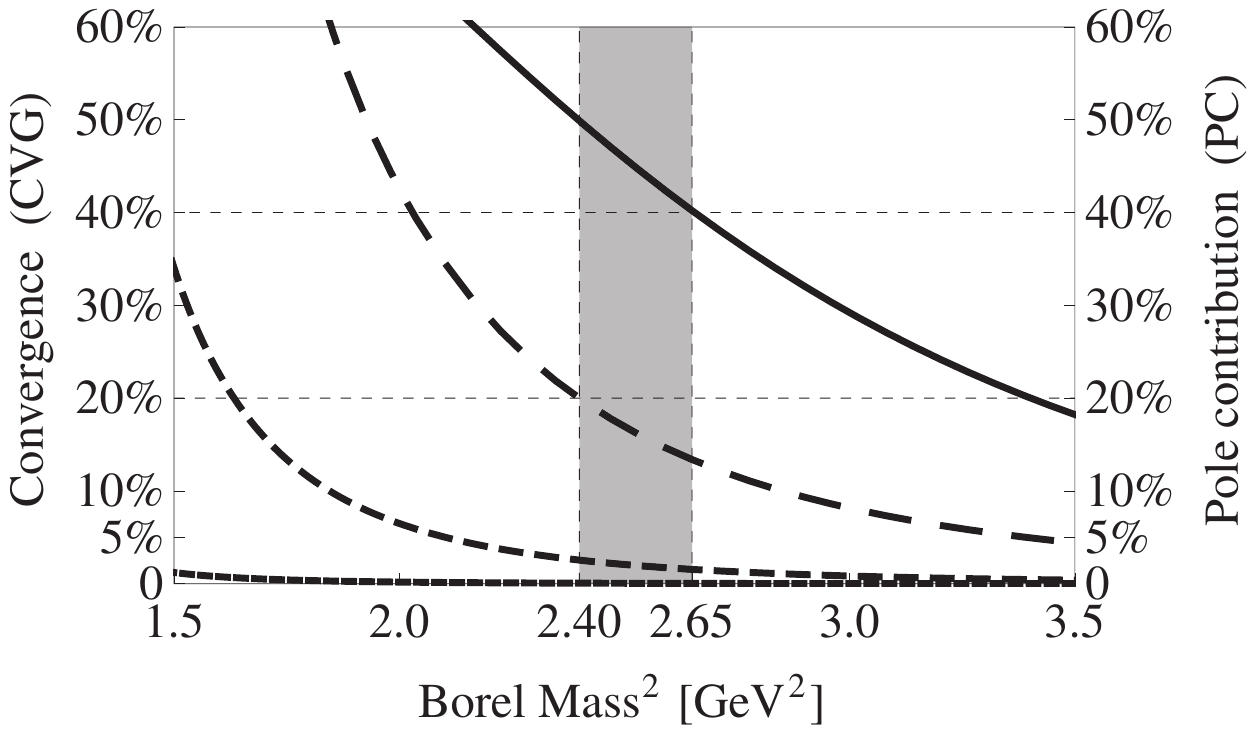}
\caption{CVG$_{12}$ (short-dashed curve, defined in Eq.~(\ref{eq:cvg12})), CVG$_{10}$ (middle-dashed curve, defined in Eq.~(\ref{eq:cvg10})), CVG$_{8}$ (long-dashed curve, defined in Eq.~(\ref{eq:cvg8})), and PC (solid curve, defined in Eq.~(\ref{eq:pc})) as functions of the Borel mass $M_B$. These curves are obtained using the current $\eta^1_{\alpha_1\alpha_2\alpha_3\alpha_4}$ when setting $s_0 = 16.0$~GeV$^2$.}
\label{fig:cvgpole}
\end{center}
\end{figure}
%%%%%%%%%%%%%%%%%%%%%%%%%%%%%%%%%%%%%%%%%%%%%%%%%%%%%%%%%%%%%%%%%%%%%%%%%%%%%%
%

Firstly, we investigate the convergence of OPE, which is the cornerstone for a reliable QCD sum rule analysis. We require the $D=12$ terms (CVG$_{12}$) to be less than 5\%, the $D=10$ terms (CVG$_{10}$) to be less than 10\%, and the $D=8$ terms (CVG$_{8}$) to be less than 20\%:
\begin{eqnarray}
\mbox{CVG}_{12} &\equiv& \left|\frac{ \Pi_{11}^{D=12}(\infty, M_B^2) }{ \Pi_{11}(\infty, M_B^2) }\right| < 5\% \, ,
\label{eq:cvg12}
\\
\mbox{CVG}_{10} &\equiv& \left|\frac{ \Pi_{11}^{D=10}(\infty, M_B^2) }{ \Pi_{11}(\infty, M_B^2) }\right| < 10\% \, ,
\label{eq:cvg10}
\\
\mbox{CVG}_8 &\equiv& \left|\frac{ \Pi_{11}^{D=8}(\infty, M_B^2) }{ \Pi_{11}(\infty, M_B^2) }\right| < 20\% \, .
\label{eq:cvg8}
\end{eqnarray}
As depicted in Fig.~\ref{fig:cvgpole} using the dashed curves, the lower bound of the Borel mass is determined to be $M_B^2 > 2.40$~GeV$^2$.

Secondly, we investigate the one-pole-dominance assumption by requiring the pole contribution (PC) to be larger than 40\%:
\begin{equation}
\mbox{PC} \equiv \left|\frac{ \Pi_{11}(s_0, M_B^2) }{ \Pi_{11}(\infty, M_B^2) }\right| > 40\% \, .
\label{eq:pc}
\end{equation}
As depicted in Fig.~\ref{fig:cvgpole} using the solid curve, the upper bound of the Borel mass is determined to be $M_B^2 < 2.65$~GeV$^2$ when setting $s_0 = 16.0$~GeV$^2$. Altogether the Borel window is determined to be $2.40$~GeV$^2 < M_B^2 < 2.65$~GeV$^2$ for $s_0 = 16.0$~GeV$^2$. Redoing the same procedures, we find that there are non-vanishing Borel windows for $s_0 > s_0^{\rm min} = 14.6$~GeV$^2$. Accordingly, we choose $s_0$ to be slightly larger, and determine its working region to be $13.0$~GeV$^2 < s_0 < 19.0$~GeV$^2$.

\begin{figure*}[hbtp]
\begin{center}
\subfigure[]{\includegraphics[width=0.4\textwidth]{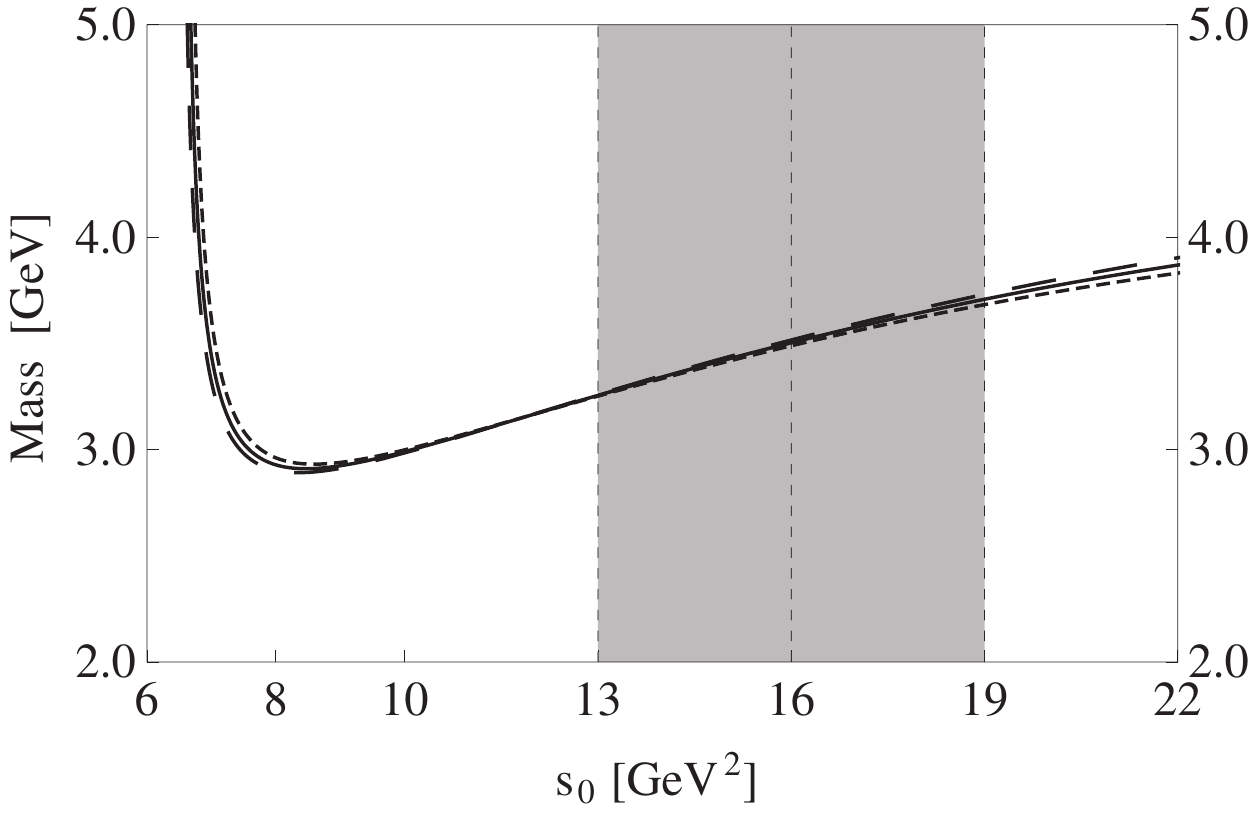}}
~~~~~~~~~~
\subfigure[]{\includegraphics[width=0.4\textwidth]{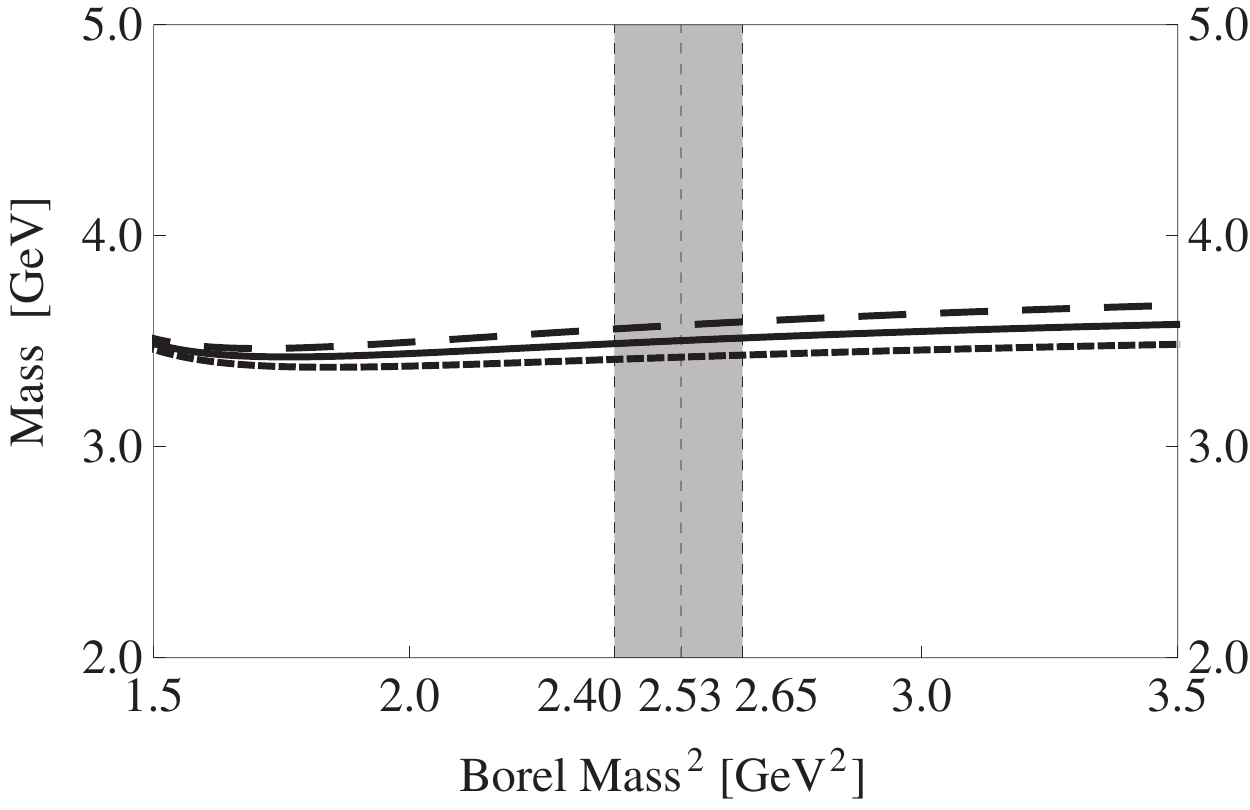}}
\caption{The mass $M_1$ of the state $X_1$ extracted from the current $\eta^1_{\alpha_1\alpha_2\alpha_3\alpha_4}$, with respect to (a) the threshold value $s_0$ and (b) the Borel mass $M_B$: (a) the short-dashed/solid/long-dashed curves are obtained by setting $M_B^2 = 2.40/2.53/2.65$~GeV$^2$, respectively; (b) the short-dashed/solid/long-dashed curves are obtained by setting $s_0 = 15.0/16.0/17.0$~GeV$^2$, respectively.}
\label{fig:mass}
\end{center}
\end{figure*}

Thirdly, we show the mass $M_1$ in Fig.~\ref{fig:mass}, and investigate its dependence on $M_B$ and $s_0$. It is stable against $M_B$ inside the Borel window $2.40$~GeV$^2 < M_B^2 < 2.65$~GeV$^2$, and its dependence on $s_0$ is moderate insider the working region $13.0$~GeV$^2 < s_0 < 19.0$~GeV$^2$, where the mass is calculated to be
\begin{equation}
M_{1} = 3.50^{+0.21}_{-0.25}{\rm~GeV} \, .
\end{equation}
Its uncertainty is due to $s_0$ and $M_B$ as well as various QCD parameters listed in Eqs.~(\ref{eq:condensates}). 

We repeat the same procedures to study the other two currents $\eta^2_{\alpha_1\alpha_2\alpha_3\alpha_4}$ and $\eta^3_{\alpha_1\alpha_2\alpha_3\alpha_4}$. The obtained results are summarized in Table~\ref{tab:results}.

\begin{table*}[hpt]
\begin{center}
\renewcommand{\arraystretch}{1.5}
\caption{QCD sum rule results for the fully-strange tetraquark states with the exotic quantum number $J^{PC} = 4^{+-}$, extracted from the diquark-antidiquark currents $\eta^{1,2,3}_{\alpha_1\alpha_2\alpha_3\alpha_4}$ as well as their mixing currents $J^{1,2,3}_{\alpha_1\alpha_2\alpha_3\alpha_4}$.}
\begin{tabular}{c|c|c|c|c|c}
\hline\hline
~~\multirow{2}{*}{Currents}~~ & ~$s_0^{min}$~ & \multicolumn{2}{c|}{Working Regions} & ~~\multirow{2}{*}{Pole~[\%]}~~ & ~~\multirow{2}{*}{Mass~[GeV]}~~
\\ \cline{3-4}
& ~~$[{\rm GeV}^2]$~~ & ~~$M_B^2~[{\rm GeV}^2]$~~ & ~~$s_0~[{\rm GeV}^2]$~~ &&
\\ \hline\hline
$\eta^1_{\alpha_1\alpha_2\alpha_3\alpha_4}$        &  14.6   &  $2.40$--$2.65$   &  $16\pm3.0$  &  $40$--$50$  &  $3.50^{+0.21}_{-0.25}$
\\
$\eta^2_{\alpha_1\alpha_2\alpha_3\alpha_4}$        &  19.2   &  $2.80$--$3.13$   &  $21\pm4.0$  &  $40$--$51$  &  $4.08^{+0.26}_{-0.31}$
\\
$\eta^3_{\alpha_1\alpha_2\alpha_3\alpha_4}$        &  11.0   &  $1.25$--$1.65$   &  $12\pm2.0$  &  $40$--$58$  &  $3.34^{+0.39}_{-0.18}$
\\ \hline
$J^1_{\alpha_1\alpha_2\alpha_3\alpha_4}$           &  10.1   &  $1.78$--$1.92$   &  $11\pm2.0$  &  $40$--$48$  &  $2.85^{+0.19}_{-0.22}$
\\
$J^2_{\alpha_1\alpha_2\alpha_3\alpha_4}$           &  19.1   &  $2.79$--$3.14$   & $21\pm4.0$   &  $40$--$51$  &  $4.08^{+0.26}_{-0.31}$
\\
$J^3_{\alpha_1\alpha_2\alpha_3\alpha_4}$           &  --     &  --               &  --          &  --          &  --
\\ \hline\hline
\end{tabular}
\label{tab:results}
\end{center}
\end{table*}

\section{Multi-channel analysis}
\label{sec:mixing}

\begin{figure}[hbt]
\begin{center}
\includegraphics[width=0.48\textwidth]{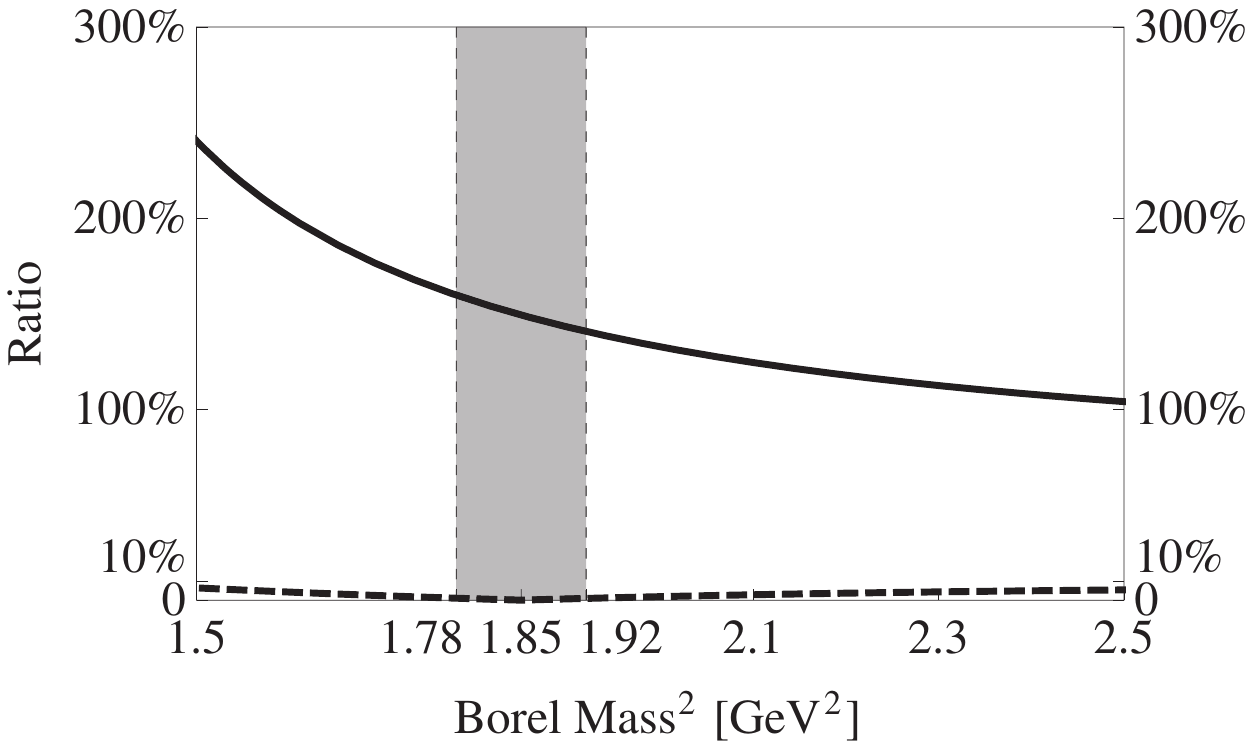}
\end{center}
\caption{Off-diagonal terms, $\left|\Pi_{13}/\sqrt{\Pi_{11}\Pi_{33}}\right|$ (solid) and $\left|\Pi^\prime_{13}/\sqrt{\Pi^\prime_{11}\Pi^\prime_{33}}\right|$ (dashed), as functions of the Borel mass $M_B$. These curves are obtained by setting $s_0 = 11.0$~GeV$^2$.}
\label{fig:offdiagonal}
\end{figure}

To perform the multi-channel analysis, we take into account the off-diagonal correlation functions, which are actually non-zero, {\it i.e.}, $\rho_{ij}(s)|_{i \neq j} \neq 0$. It is interesting to see how large they are, so we choose $s_0 = 11.0$~GeV$^2$ and $M_B^2 = 1.85$~GeV$^2$ to obtain
\begin{equation}
\Pi_{ij}(s_0, M_B^2)
=
\left(\begin{array}{ccc}
2.77 & -0.04 & -3.83
\\
-0.04 & 0.98 & 0.46
\\
-3.83 & 0.46 & 2.38
\end{array}\right) \times 10^{-6} {\rm~GeV}^{14} .
\end{equation}
Hence, $\eta^{1}_{\cdots}$ and $\eta^{3}_{\cdots}$ are strongly correlated with each other, making the off-diagonal terms of $\rho_{ij}(s)$ non-negligible, as depicted in Fig.~\ref{fig:offdiagonal} using the solid curve.

To diagonalize the $3\times3$ matrix $\rho_{ij}(s)$, we construct three mixing currents $J^{1,2,3}_{\alpha_1\alpha_2\alpha_3\alpha_4}$:
\begin{equation}
\left(\begin{array}{c}
J^{1}_{\cdots}
\\
J^{2}_{\cdots}
\\
J^{3}_{\cdots}
\end{array}\right)
=
\mathbb{T}_{3\times3}
\left(\begin{array}{c}
\eta^{1}_{\cdots}
\\
\eta^{2}_{\cdots}
\\
\eta^{3}_{\cdots}
\end{array}\right)
\, ,
\label{eq:transition}
\end{equation}
with $\mathbb{T}_{3\times3}$ the transition matrix.

We apply the method of operator product expansion to extract the spectral densities $\rho^\prime_{ij}(s)$ from the mixing currents $J^{1,2,3}_{\alpha_1\alpha_2\alpha_3\alpha_4}$. After choosing
\begin{equation}
\mathbb{T}_{3\times3}
=
\left(\begin{array}{ccc}
0.72 & -0.06 & -0.69
\\
0.14 & 0.99 & 0.05
\\
0.68 & -0.13 & 0.72
\end{array}\right) \, ,
\label{eq:matrix}
\end{equation}
we obtain
\begin{equation}
\Pi^\prime_{ij}(s_0, M_B^2)
=
\left(\begin{array}{ccc}
6.43 & 0 & 0
\\
0 & 1.00 & 0
\\
0 & 0 & -1.30
\end{array}\right) \times 10^{-6} {\rm~GeV}^{14} ,
\label{eq:mixingpi}
\end{equation}
at $s_0 = 11.0$~GeV$^2$ and $M_B^2 = 1.85$~GeV$^2$. Hence, the off-diagonal terms of $\rho^\prime_{ij}(s)$ are negligible around here, suggesting that the three mixing currents $J^{1,2,3}_{\alpha_1\alpha_2\alpha_3\alpha_4}$ are nearly non-correlated around here, as depicted in Fig.~\ref{fig:offdiagonal} using the dashed curve. Moreover, Eq.~(\ref{eq:mixingpi}) indicates that the QCD sum rule result from $J^3_{\alpha_1\alpha_2\alpha_3\alpha_4}$ is non-physical around here due to its negative correlation function. Besides, Eq.~(\ref{eq:matrix}) indicates that $J^{2}_{\cdots}$ is almost the same as $\eta^{2}_{\cdots}$, while $J^{1}_{\cdots}$ and $J^{3}_{\cdots}$ are mainly from the recombination of $\eta^{1}_{\cdots}$ and $\eta^{3}_{\cdots}$.

\begin{figure*}[hbtp]
\begin{center}
\subfigure[]{\includegraphics[width=0.4\textwidth]{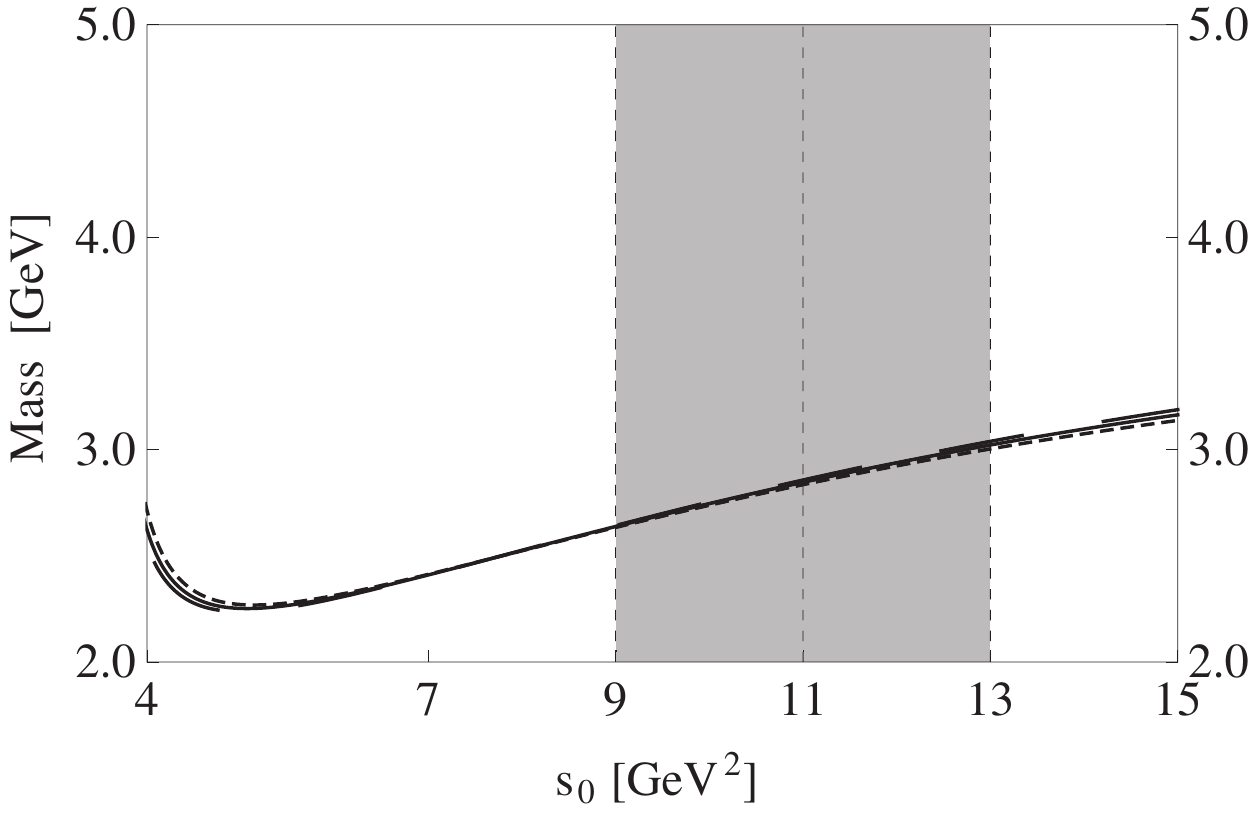}}
~~~~~~~~~~
\subfigure[]{\includegraphics[width=0.4\textwidth]{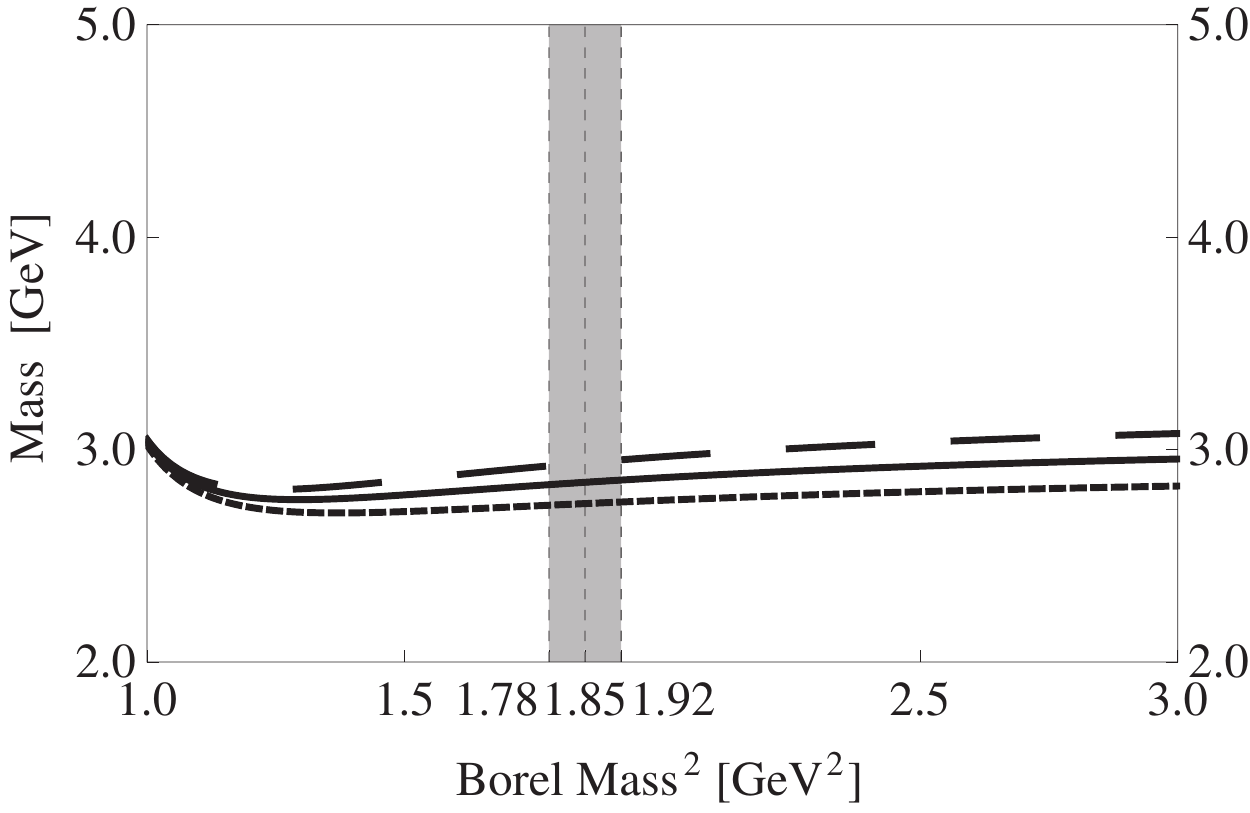}}
\caption{The mass $M_1^\prime$ extracted from the mixing current $J^1_{\alpha_1\alpha_2\alpha_3\alpha_4}$, with respect to (a) the threshold value $s_0$ and (b) the Borel mass $M_B$: (a) the short-dashed/solid/long-dashed curves are obtained by setting $M_B^2 = 1.78/1.85/1.92$~GeV$^2$, respectively; (b) the short-dashed/solid/long-dashed curves are obtained by setting $s_0 = 10.0/11.0/12.0$~GeV$^2$, respectively.}
\label{fig:massmix}
\end{center}
\end{figure*}

We use the procedures previously applied on the diquark-antidiquark currents $\eta^{1,2,3}_{\alpha_1\alpha_2\alpha_3\alpha_4}$ to study their mixing currents $J^{1,2,3}_{\alpha_1\alpha_2\alpha_3\alpha_4}$. The obtained results are also summarized in Table~\ref{tab:results}. Especially, the mass extracted from the current $J^1_{\alpha_1\alpha_2\alpha_3\alpha_4}$ is significantly reduced to be
\begin{equation}
M^\prime_{1} = 2.85^{+0.19}_{-0.22} {\rm~GeV} \, .
\end{equation}
For completeness, we show it in Fig.~\ref{fig:massmix} as a function of the threshold value $s_0$ and the Borel mass $M_B$.

\section{Conclusion}
\label{sec:summary}

In this paper we apply the method of QCD sum rules to study the fully-strange tetraquark states with the exotic quantum number $J^{PC}=4^{+-}$. We work within the diquark-antidiquark picture and systematically construct their interpolating currents. We calculate both the diagonal and off-diagonal correlation functions. The obtained results are used to construct three mixing currents that are nearly non-correlated. We use the mixing current $J^1_{\alpha_1\alpha_2\alpha_3\alpha_4}$ to evaluate the mass of the lowest-lying state to be $2.85^{+0.19}_{-0.22}$~GeV.

In this paper we also systematically construct the fully-strange meson-meson currents of $J^{PC}=4^{+-}$, and relate them to the diquark-antidiquark currents through the Fierz rearrangement. Especially, we can apply Eq.~(\ref{eq:transition}) and Eq.~(\ref{eq:fierz}) to transform the mixing current $J^1_{\alpha_1\alpha_2\alpha_3\alpha_4}$ to be
\begin{equation}
J^1_{\cdots} = 4.3~\xi^1_{\cdots} + 1.2~\xi^2_{\cdots} - 1.3~\xi^3_{\cdots} \, .
\end{equation}
This Fierz identity suggests that the lowest-lying state dominantly decays into the $P$-wave $\phi(1020) f_2^\prime(1525)$ channel through the meson-meson current $\xi^1_{\alpha_1\alpha_2\alpha_3\alpha_4}$, given that the operator $\bar{s}_b \gamma_{\alpha_2} s_b$ of $I^GJ^{PC}=0^-1^{--}$ well couples to the vector meson $\phi(1020)$ and the operator $\mathcal{S}[\bar s_a \gamma_{\alpha_1} {\overset{\leftrightarrow}{D}}_{\alpha_3} s_a]$ of $I^GJ^{PC}=0^+2^{++}$ well couples to the $f_2^\prime(1525)$ meson. Accordingly, we propose to search for it in the $X \to \phi(1020) f_2^\prime(1525) \to \phi K \bar K$ decay process in the future Belle-II, BESIII, COMPASS, GlueX, and PANDA experiments.

This is the first study on the exotic quantum number $J^{PC}=4^{+-}$, and the above lowest-lying fully-strange tetraquark state of $J^{PC}=4^{+-}$ is a purely exotic hadron to be potentially observed in future experiments. Its theoretical and experimental studies will continuously improve our understanding on the non-perturbative behaviors of the strong interaction in the low energy region.

\begin{acknowledgements}
We thank Wei Chen, Er-Liang Cui, and Hui-Min Yang for useful discussions.
This project is supported by
the National Natural Science Foundation of China under Grant No.~12075019,
and
the Fundamental Research Funds for the Central Universities.
\end{acknowledgements}

\end{document}